\documentclass{wscpaperproc}
\usepackage{latexsym}
\usepackage{graphicx}
\usepackage{mathptmx}
\usepackage[T1]{fontenc}

\input{definitions}

\usepackage{amsmath}
\usepackage{amsfonts}
\usepackage{amssymb}
\usepackage{amsbsy}
\usepackage{amsthm}

\usepackage[pdftex,colorlinks=true,urlcolor=blue,citecolor=black,anchorcolor=black,linkcolor=black]{hyperref}

\newtheoremstyle{wsc}
{3pt}
{3pt}
{}
{}
{\bf}
{}
{.5em}
{}

\theoremstyle{wsc}

\begin{document}

%
%

\pagestyle{fancyplain}

\thispagestyle{plain}
\firstPageHead{}

\chead{\fancyplain{}{\itshape Id\'e and Miyaguchi}}

\rhead{}
\cfoot{}
\renewcommand{\headrulewidth}{0pt} 

\makeatletter
\let\@internalcite\cite
\def\cite{\def\@citeseppen{-1000}%
    \def\@cite##1##2{(##1\if@tempswa , ##2\fi)}%
    \def\citeauthoryear##1##2##3{##1 ##3}\@internalcite}
\def\citeNP{\def\@citeseppen{-1000}%
    \def\@cite##1##2{##1\if@tempswa , ##2\fi}%
    \def\citeauthoryear##1##2##3{##1 ##3}\@internalcite}
\def\citeN{\def\@citeseppen{-1000}%
    \def\@cite##1##2{##1\if@tempswa, ##2)\else{}\fi}%
    \def\citeauthoryear##1##2##3{##1 (##3)}\@citedata}
\def\citeA{\def\@citeseppen{-1000}%
    \def\@cite##1##2{(##1\if@tempswa , ##2\fi)}%
    \def\citeauthoryear##1##2##3{##1}\@internalcite}
\def\citeANP{\def\@citeseppen{-1000}%
    \def\@cite##1##2{##1\if@tempswa , ##2\fi}%
    \def\citeauthoryear##1##2##3{##1}\@internalcite}
\def\shortcite{\def\@citeseppen{-1000}%
    \def\@cite##1##2{(##1\if@tempswa , ##2\fi)}%
    \def\citeauthoryear##1##2##3{##2 ##3}\@internalcite}
\def\shortciteNP{\def\@citeseppen{-1000}%
    \def\@cite##1##2{##1\if@tempswa , ##2\fi}%
    \def\citeauthoryear##1##2##3{##2 ##3}\@internalcite}
\def\shortciteN{\def\@citeseppen{-1000}%
    \def\@cite##1##2{##1\if@tempswa, ##2\else{}\fi}%
    \def\citeauthoryear##1##2##3{##2 (##3)}\@citedata}
\def\shortciteA{\def\@citeseppen{-1000}%
    \def\@cite##1##2{(##1\if@tempswa , ##2\fi)}%
    \def\citeauthoryear##1##2##3{##2}\@internalcite}
\def\shortciteANP{\def\@citeseppen{-1000}%
    \def\@cite##1##2{##1\if@tempswa , ##2\fi}%
    \def\citeauthoryear##1##2##3{##2}\@internalcite}
\def\citeyear{\def\@citeseppen{-1000}%
    \def\@cite##1##2{(##1\if@tempswa , ##2\fi)}%
    \def\citeauthoryear##1##2##3{##3}\@citedata}
\def\citeyearNP{\def\@citeseppen{-1000}%
    \def\@cite##1##2{##1\if@tempswa , ##2\fi}%
    \def\citeauthoryear##1##2##3{##3}\@citedata}
%
%
%
\def\@citedata{%
    \@ifnextchar [{\@tempswatrue\@citedatax}%
                  {\@tempswafalse\@citedatax[]}%
}

\def\@citedatax[#1]#2{%
\if@filesw\immediate\write\@auxout{\string\citation{#2}}\fi%
  \def\@citea{}\@cite{\@for\@citeb:=#2\do%
    {\@citea\def\@citea{, }\@ifundefined
       {b@\@citeb}{{\bf ?}%
       \@warning{Citation `\@citeb' on page \thepage \space undefined}}%
{\csname b@\@citeb\endcsname}}}{#1}}%

%
\def\@citex[#1]#2{%
\if@filesw\immediate\write\@auxout{\string\citation{#2}}\fi%
  \def\@citea{}\@cite{\@for\@citeb:=#2\do%
    {\@citea\def\@citea{; }\@ifundefined
       {b@\@citeb}{{\bf ?}%
       \@warning{Citation `\@citeb' on page \thepage \space undefined}}%
{\csname b@\@citeb\endcsname}}}{#1}}%

%
\def\@biblabel#1{}
\makeatother



\newdimen\bibindent
\bibindent=0.0em
\def\thebibliography#1{\section*{\refname}\list
   {}{\settowidth\labelwidth{[#1]}
   \leftmargin\parindent
   \itemindent -\parindent
   \listparindent \itemindent
   \itemsep 0pt
   \parsep 0pt}
   \def\newblock{}
   \sloppy
   \sfcode`\.=1000\relax}


\setlength{\baselineskip}{12.7pt}

\title{Cross-Process Defect Attribution using Potential Loss Analysis}

\author{\begin{center}Tsuyoshi Id\'{e}\textsuperscript{1}, Kohei Miyaguchi\textsuperscript{2}\thanks{Kohei Miyaguchi is currently affiliated with LY Research, Japan.}\\
[11pt]
\textsuperscript{1}IBM Semiconductors, IBM Thomas J. Watson Research Center, New York, USA.\\
\textsuperscript{2}IBM Research -- Tokyo, Tokyo, Japan.\end{center}
}

\maketitle

\vspace{-12pt}

\section*{ABSTRACT}
Cross-process root-cause analysis of wafer defects is among the most critical yet challenging tasks in semiconductor manufacturing due to the heterogeneity and combinatorial nature of processes along the processing route. 
This paper presents a new framework for wafer defect root cause analysis, called Potential Loss Analysis (PLA), as a significant enhancement of the previously proposed partial trajectory regression approach. The PLA framework attributes observed high wafer defect densities to upstream processes by comparing the best possible outcomes generated by partial processing trajectories. 
We show that the task of identifying the best possible outcome can be reduced to solving a Bellman equation. Remarkably, the proposed framework can simultaneously solve the prediction problem for defect density as well as the attribution problem for defect scores. 
We demonstrate the effectiveness of the proposed framework using real wafer history data.

\section{Introduction}

The latest technology nodes in semiconductor manufacturing involve more than one thousand process steps across about a dozen process types such as deposition and etching. Cross-process root-cause analysis of wafer defects spanning the entire processing sequence is among the most critical yet challenging tasks in semiconductor manufacturing. Particularly during process integration and yield ramp-up stages, classical design-of-experiment methodologies, which analyze process outcomes across systematically varied parameters, are often impractical due to the prohibitively large number of adjustable parameters along the processing route. Although a wide range of off-the-shelf machine learning tools are publicly available, most of these tools are designed for prediction tasks, such as estimating real-valued outputs (regression) or categorical outcomes (classification). As a result, fab-wide defect diagnosis still heavily depends on manual, ad hoc analysis by domain experts. 

To support more systematic analysis, three major directions have been pursued in the semiconductor analytics literature to date. The first approach treats defect diagnosis as a by-product of cross-process virtual metrology (VM) modeling. Regularized linear regression combined with variable selection techniques is commonly used (e.g.,~\cite{susto2015multi,jebri2016virtual,kim2018variable}). However, linear models struggle to capture complex nonlinear relationships across heterogeneous fabrication processes. Furthermore, defect attribution based on linear models is essentially reduced to variable-wise correlation analysis, which is known to yield only weak attribution signals~\cite{Miyaguchi25ASMC}. 

The second direction is to leverage recurrent neural networks (RNNs) and Transformers to replace conventional VM models. While they can capture complex nonlinear dependencies in sequential processes (e.g.,~\cite{yella2021soft,han2023deep,dalla2023deep,lee2020recurrent,hsu2023virtual}), handling different processes requires significant feature engineering effort. Additionally, they typically operate as black boxes, making input attribution a non-trivial task. These models are also generally data-intensive, and it is often difficult to collect enough data to cover the combinatorial complexity of the fabrication process. 

The third direction involves leveraging explainable artificial intelligence (XAI) techniques applicable to black-box prediction models. This is a promising direction in that it can potentially enhance expressive prediction models with interpretability, helping to identify which process steps should be adjusted to improve defect rates. However, for cross-process defect attribution, existing methods, such as those based on Shapley values (e.g.,~\cite{torres2020machine,senoner2022using,lee2023expandable,guo2024enhanced}), pay limited attention to key characteristics of semiconductor processes, such as the sequential nature of fabrication steps. In addition, they often rely on assumptions that may not be fully justifiable in semiconductor manufacturing, such as dependence on arbitrarily selected baseline inputs. Ironically, these XAI approaches are often used as yet another form of black-box reasoning without careful justifications. 

\begin{figure}[t]
    \centering
    \includegraphics[clip, trim=0cm 7.5cm 0cm 3cm, width=0.9\linewidth]{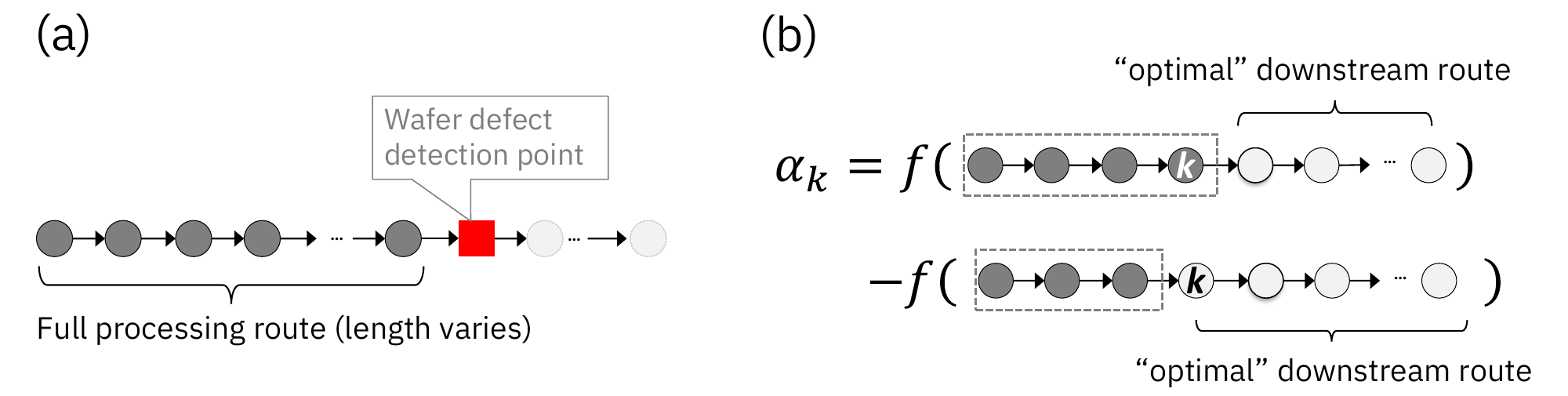}
    \vspace{-2mm}
    \caption{Problem setting and the key idea. (a) We are interested in identifying upstream processes responsible for wafer defects detected at a specific detection point. The number of process steps along the route can vary. (b) The key idea of the PLA approach. Instead of zeroing out the process embeddings on the downstream path, it solves an optimization problem for optimal downstream routes.}
    \label{fig:PTR_concept}
\end{figure}

Recently, a new framework called the partial trajectory regression (PTR) was proposed to address these issues, such as capturing the sequential nature of fabrication and the lack of direct attribution capability~\cite{Miyaguchi25PTR}. However, similar to Shapley-value-based approaches, its attribution mechanism still suffers from potential biases due to inappropriate model assumptions, as discussed in detail later.

In this paper, we propose a novel framework called \textit{potential loss analysis} (PLA), as a significant enhancement over PTR. Figure~\ref{fig:PTR_concept} illustrates the key idea. Our goal is to attribute an observed wafer quality issue by computing a responsibility score (or attribution score) for each upstream process. To evaluate the influence of the $k$-th process, we compare counterfactual outcomes based on partial process routes with and without the target process. The key idea is to use \textit{optimal} downstream routes and compare the outcomes from the best possible continuations. To identify such optimal routes, we formulate wafer processing as a sequential decision-making problem and solve a Bellman equation. To the best of our knowledge, this is the first work to introduce the notion of path optimization into wafer defect attribution. We demonstrate the effectiveness of the proposed framework using real wafer history data from a state-of-the-art FEOL (front-end-of-line) process.

\section{Related Work}

This section provides more detailed context of the problem we address with a particular focus on cross-process virtual metrology and explainable AI.

\subsection{Cross-Process Virtual Metrology}

A key characteristic that distinguishes semiconductor manufacturing from other industrial domains is its process complexity. A typical semiconductor process involves more than hundreds of intricate and highly specialized physical operations, including photolithography, thermal annealing, polishing, wet and dry etching, ion implantation, electroplating, sputtering, and chemical vapor deposition, among others.

For cross-process root-cause analysis, these heterogeneous operations must be mapped to a shared representation space to enable meaningful comparisons. Existing literature offers three general approaches for this. 
The first approach utilizes process trace data~\cite{xu2024fast,fan2022key}. While effective within individual process tools, this method requires extensive tool-specific preprocessing, and the quality of analysis heavily depends on the chosen preprocessing strategy, making it less suitable for cross-process analysis. 
The second approach uses inline measurements as proxies for physical processes. Since these measurements partially absorb the physical heterogeneity across processes, this method has become common practice in recent studies~\cite{senoner2022using,guo2024enhanced,wang2024improved,ni2025novel}. However, these approaches typically disregard the sequential order of processes and perform root-cause analysis as a by-product of virtual metrology (VM), often via univariate correlation analysis, which is known to yield only weak attribution signals~\cite{Miyaguchi25ASMC}.

The third approach involves embedding techniques, where data objects (e.g., process steps) are transformed into numerical vector representations. For instance, Fan et al.~\cite{fan2022data} use one-hot encoding to unify categorical and numerical data in the VM setting. Schulz et al.~\cite{schulz2022graph} propose defining a fab state vector using known interdependencies among processing tools under an unsupervised setting, without the context of wafer defect analysis. More recently, Miyaguchi et al.~\cite{Miyaguchi25PTR} proposed proc2vec and route2vec algorithms that think of process attributes as synthetic words and capture their similarity using kernel embedding. The proposed PLA framework uses their approach as a building block.

\subsection{Explainable AI (XAI)}

Numerous methods have been developed to improve the interpretability of machine learning models under the umbrella of XAI~\cite{xu2019explainable}. One widely used category is additive explanation methods~\cite{lundberg2017unified,ribeiro2016why}, which provide mathematically justified decompositions of a model's output into individual contributions of input variables. Other common XAI techniques include gradient-based methods~\cite{selvaraju2020grad} and attention-based methods~\cite{ali2022xai}. 

While most XAI studies have traditionally assumed vector inputs whose dimensions can be arbitrarily reordered, a growing—though still relatively limited—body of research is beginning to address the unique challenges posed by sequential data, particularly time-series data~\cite{rojat2021explainable}. In the specific field of semiconductor analytics, the integration of model-agnostic XAI techniques with advanced VM models is emerging as a research trend, aiming to balance predictive power with interpretability. Among the wide variety of XAI methods (see, e.g.,\cite{molnar2020interpretable} for an overview), the majority of recent studies adopt the Shapley value~\cite{torres2020machine,senoner2022using,lee2023expandable,guo2024enhanced}, possibly due to the availability of a well-designed Python implementation~\cite{lundberg2017unified}. Ironically, despite its widespread adoption, most studies apply XAI methods as black boxes, without critically examining their modeling assumptions. In fact, mainstream attribution algorithms, such as Shapley values and integrated gradients, provide attribution scores \textit{relative to} an arbitrary reference point. A similar issue arises in the PTR framework, as discussed later. 

In contrast, the proposed PLA framework, which is based on our recent unpublished work~\cite{Miyaguchi25advantage_RL}, completely eliminates the need for the arbitrary reference point, which we believe presents a major step forward in XAI research.

\section{Preliminaries}

This section provides a formal definition of the attribution problem and an overview of partial trajectory regression as the baseline approach. 

\subsection{Problem Setting}

Our main goal is to develop a method for computing the attribution score of each process in a wafer's processing route, given an observed process outcome. To formulate the attribution model, we assume a training dataset consisting of $N$ pairs of (process outcome metric, processing route): 
\begin{align}\label{eq:trajectory_as_(x,t)sequence}
    \calD \triangleq \{ (y^{(n)}, \xi^{(n)})  \mid n=1,\ldots, N\}, \quad \xi^{(n)} = \left((\bmx_1^{(n)}, t_1^{(n)}),\ldots, (\bmx_{L^{(n)}}^{(n)}, t_{L^{(n)}}^{(n)})\right),
\end{align}
where $N$ is the number of wafers and the superscript $^{(n)}$ indicates that the quantity belongs to the $n$-th wafer. The symbols $\xi$ and $y$ denote a processing route (or trajectory) and the corresponding process outcome metric, respectively. For $y$, we use log defect density in empirical evaluations. $L$ denotes the number of processes in route $\xi$. Note that $L$ may vary across wafers due to reworks or different route definitions. Therefore, treating $\xi$ as a fixed-dimensional object is generally not appropriate. 

We assume that process $k$ ($k=1,\ldots,L)$ has a vector representation $\bmx_k \in \mathbb{R}^D$ and an associated timestamp~$t_k$, where $D$ is the dimensionality of the representation space. While finding a common representation space across all the processes is a nontrivial task, the kernel embedding method described in the next subsection provides a practical solution. 

Because the length of each trajectory varies, the index $k$ only refers to the position of a process within a given trajectory $\xi$. For instance, the 10th process in $\xi^{(1)}$ and the 10th process in $\xi^{(2)}$ may correspond to entirely different physical operations.

Formally, the task of process attribution is defined as follows:
\begin{definition}[Process attribution]
    Find a function $\alpha_k(\xi)$ that computes the responsibility score for the $k$-th process on a process route instance $\xi$ ($k=1,\ldots, L$), which is generally not included in $\calD$.
\end{definition}

\subsection{Process Embedding}

We assume that processes are represented by numerical vectors $\{\bmx_k\}_{k=1}^L$. As discussed in the previous section, this is a nontrivial task due to the heterogeneity of semiconductor processes, which include etching, polishing, ion implantation, and more. Here, we adopt the kernel embedding approach proposed by~\cite{Miyaguchi25PTR}, assuming that high-level process attributes such as process ID and recipe ID are available from the manufacturing execution system (MES). 

We first extract common attributes from the MES across different processes, such as equipment IDs, recipe IDs, tool types, photo layer IDs, route IDs, and others. We then create a synthetic `token' for each process by concatenating these strings as follows:
\begin{align}\nonumber
    (\text{process token}) = \texttt{eqp} \oplus
    \texttt{recipe} \oplus
    \texttt{tool\_type} \oplus \texttt{photo\_layer} \oplus
    \texttt{route} \oplus \cdots,
\end{align}
where $\oplus$ denotes string concatenation with a suitable separator.

Based on these string representations, we construct a dictionary of tokens. Let $V_d$ denote the size of the vocabulary, i.e.,~the number of unique tokens. For this dictionary, we compute a kernel matrix $\sfK \in \mathbb{R}^{V_d \times V_d}$, where $K_{i,j}$ is the similarity between tokens $i$ and $j$ computed using a variant of the substring kernel~\cite{lodhi2002text,shawe2004kernel}. Once the kernel matrix $\sfK$ is computed, the vector representation of token $i$ is obtained as:
\begin{align}\label{eq:embedding_MDS}
    \bmx_i = (\sqrt{\lambda}_1v_{1,i}, \ldots, \sqrt{\lambda}_k v_{k,i}, \ldots, \sqrt{\lambda}_D v_{D,i})^\top,
\end{align}
where $\lambda_k$ is the $k$-th largest eigenvalue of $\sfK$, and $v_{k,i}$ is the $i$-th element of the eigenvector corresponding to $\lambda_k$. $D$ is a user-defined embedding dimensionality.

\subsection{Partial Trajectory Regression}
\label{subsec:PTR}

With the process embeddings $\{\bmx_k\}$ now represented in a $D$-dimensional space, we aim to build a predictive model for defect density $y$ as a function of the process vector sequence. This remains a nontrivial task due to the variable length of trajectories, making it an instance of \textit{trajectory regression}~\cite{ide2009travel,ide2011trajectory}, distinct from standard regression problems.

\begin{figure}[tb]
    \centering
    \includegraphics[clip, trim=0cm 10cm 16cm 4cm, width=0.65\linewidth]{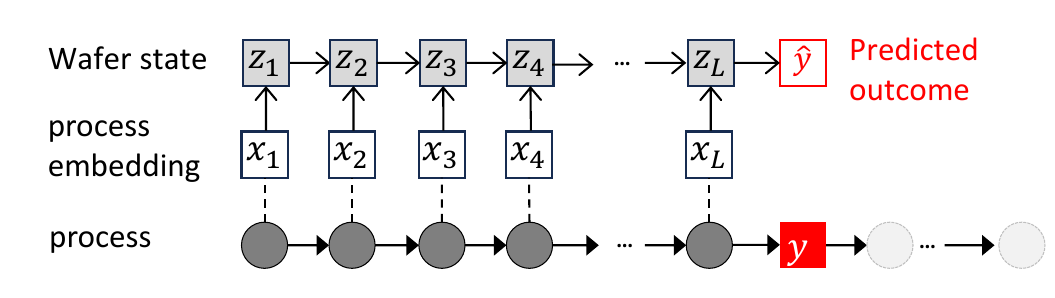}
    \caption{Assumed state-space model of PLA. Each process has a vector representation called the process embedding. The wafer state evolves based on the process embedding and the previous state.}
    \label{fig:PTR_architecture}
\end{figure}

In this work, we employ a state-space model as illustrated in Fig.~\ref{fig:PTR_architecture}. A general form of the model is defined by an RNN architecture:
\begin{align}\label{eq:cell_state_transition}
    \bmx_k = \text{Embed}(\text{process}_k), \quad 
    \bmz_k = \text{Cell}(\bmz_{k-1}, \bmx_k, t_k),
\end{align}
where $\bmz_k$ denotes the wafer state vector after applying the first $k$ processes, and $\bmx_k$ is the process embedding from the previous section. $t_k$ denotes the timestamp associated with $\bmx_k$. The function $\text{Cell}(\cdot)$ is typically implemented as a neural network that computes a new wafer state from the previous state and current input $(\bmx_k, t_k)$. While standard RNN cells such as Long Short-Term Memory (LSTM) or Gated Recurrent Unit (GRU) can be used, it has been suggested that a custom lightweight architecture
\begin{align}\label{eq:cell_linear}
    \bmz_k = \psi(t_k, t_{k-1}) \bmx_k + \bmz_{k-1},
\end{align}
is preferable in wafer root-cause analysis scenarios due to the lack of sufficient wafer samples covering the entire process variability~\cite{Miyaguchi25PTR}. Here, $\psi(\cdot,\cdot)$ is a temporal weighting function, set to $\log_{10}(1 + \cdot)$ in our experiments. 

One notable property of the PTR architecture in Fig.~\ref{fig:PTR_architecture} is its ability to make predictions based on \textit{partial trajectories}. This is achieved by training a regression model of the form:
\begin{align}\label{eq:regression_MLP}
    f_{\phi}(\bmz_k) = \text{MLP}_{\phi}(\bmz_k),
\end{align}
where $\text{MLP}_{\phi}$ is a multi-layer perceptron with parameters $\phi$, trained to minimize the loss:
\begin{align}
    L_\text{PTR}(\phi) \triangleq \frac{1}{2N} \sum_{n=1}^N \sum_{k=1}^{L^{(n)}} \frac{1}{L^{(n)}} \left(y^{(n)} - f_\phi(\bmz_k)\right)^2 + \eta \|\phi\|_1,
\end{align}
where $\|\cdot\|_1$ denotes the $\ell_1$ norm and $\eta$ is its regularization strength as a hyperparameter.

\section{Potential Loss Analysis Framework}

This section begins by discussing the limitations of the PTR-based attribution strategy and proceeds to introduce the proposed optimized potential outcome framework. 

\subsection{Issues with PTR-based Attribution}

The PTR framework provides a useful approach to cross-process defect attribution. As illustrated in Fig.~\ref{fig:PTR_concept}, the PTR attribution method computes the attribution score of process $k$ by comparing two potential outcomes defined by partial trajectories. Specifically, the attribution score of the $k$-th process in $\xi$ is computed as:
\begin{align}\label{eq:PTR_attribution}
    \alpha_k(\xi)=
        f(\bmz_k)-f(\bmz_{k-1}), \quad \quad \text{(PTR)}
\end{align}
where $f(\cdot)$ is a regression function used to predict the process outcome, as discussed in Sec.~\ref{subsec:PTR}. Unless $f(\bmz_k)$ is an additive function over different $k$s, the score $\alpha_k(\xi)$ should include not only single-process effects but also multi-process correlations with upstream processes. This approach aligns with Rubin's potential outcome framework~\cite{rubin2005causal}, as it simulates a counterfactual comparison with and without the target process $k$, similar to the situation in randomized controlled trials.

A critical question, however, is whether the PTR attribution method~\eqref{eq:PTR_attribution} is able to isolate the effect of process $k$ alone, independent of all other factors. Notably, one can show that the prediction on a partial trajectory is equivalent to that of the full trajectory with the downstream process vectors set to $\bmzero$. This follows from the equivalent form of Eq.~\eqref{eq:cell_linear}:
\begin{align}
    \bmz_k = \bmz_0 + \sum_{i=1}^k \psi(t_i, t_{i-1}) \bmx_i.
\end{align}
Thus, the full trajectory $\bmz_L$ equals a partial trajectory $ \bmz_k$ if $\bmx_{k+1} = \cdots = \bmx_L = \bmzero$.

Since the scale and origin of the process vectors can be arbitrary, this "zeroing out" approach may introduce biases into the attribution score. This is true, for example, when a dimension of $\bmx_i$ represents a binary indicator variable. While process-wise standardization might help mitigate this issue, such an approach results in unwanted dependency of the attribution score on the population means, depending on the prediction model adopted, as shown in Sec.~\ref{sec:experiments} (see Fig.~\ref{fig:CT_scoring}). This is the same type of issue encountered with Shapley values, whose attribution scores are defined relative to an arbitrary reference point. We next discuss how to eliminate this arbitrariness.

\subsection{Defining Optimal Expected Cumulative Loss}

The key idea of the proposed Potential Loss Analysis (PLA) framework is to use an optimized downstream route rather than arbitrarily zeroing out the process embeddings.

Let $F(\bmz_k \mid \bmx_{k+1}, \ldots, \bmx_L)$ denote the predicted outcome starting from wafer state $\bmz_k$ when the wafer follows a downstream path $\bmx_{k+1}, \ldots, \bmx_L$ afterwards. We model this as a sequential decision-making problem, where the next process $\bmx_{k+1}$ is chosen based on $\bmz_k$. Each action transitions the wafer state to a new state $\bmz_{k+1}$ and incurs a loss $C(\bmz_{k+1})$. We define the cumulative expected loss from state $\bmz_1$ as:
\begin{align}\label{eq:F(z|xxxx)}
    F(\bmz_1 \mid \bmx_1, \bmx_2, \ldots) = \mathbb{E}\left[ 
    \sum_{t=1}^\infty C(\bmz_t)
    \mid \bmz \right],
\end{align}
where we used the numbering from one as the subscripts specify the relative positions within a trajectory. Each transition $(\bmz,\bmx)\to \bmz'$ may be affected by random factors. We model such randomness using a probability distribution $p(\bmz' \mid \bmz, \bmx)$ and the expectation $\mathbb{E}[\cdot]$ is taken with respect to this distribution. We assume that the transition terminates at $\bmz_L \in \calS_T$, where $\calS_T$ denotes the set of terminal states. The wafer state remains in an absorbing state $\bms_\perp$ after the terminal state:
\begin{align}\label{eq:absorbing_transition}
    p(\bmz' \mid \bmz,\bmx) = \delta(\bmz' - \bms_\perp) \quad 
    \text{for } \bmz \in \calS_T \text{ or } \bmz = \bms_\perp,
\end{align}
where $\delta(\cdot)$ is the Dirac delta function.

In our problem setting, the instantaneous loss function $C(\bmz)$ represents the defect density observed at the terminal state. Formally, it is given by:
\begin{align}\label{eq:terminal_loss_def}
    C(\bmz) =\begin{cases}
        y(\bmz), &  \bmz \in \calS_T \\        
        0, & \text{otherwise},
    \end{cases}
\end{align}
where $y(\bmz)$ is the observable defect density at the terminal state. From Eqs.~\eqref{eq:F(z|xxxx)} and~\eqref{eq:absorbing_transition}, we have
\begin{align}\label{eq:F=0_at_absorbed state}
F(\bmz \in \calS_T \mid \bmx_1,\bmx_2,\ldots) = \mathbb{E}\left[y(\bmz)\right] 
\quad \text{and} \quad
    F(\bmz=\bmz_\perp\mid \bmx_1,\bmx_2,\ldots) = \mathbb{E}\left[0+0+\cdots \right] = 0.
\end{align}

We now define the \textit{optimal expected cumulative loss}, which serves as a replacement for $f(\cdot)$ in Eq.~\eqref{eq:PTR_attribution}:
\begin{align}\label{eq:optimal_expected_cumulative_loss_def1}
    F^*(\bmz_1) &\triangleq \min_{\bmx_1,\bmx_2,\ldots} F(\bmz_1 \mid \bmx_1, \bmx_2, \ldots)
    = \min_{\bmx_1} \left\{ 
    C(\bmz_1) + \sum_{\bmz_2} p(\bmz_2 \mid \bmz_1, \bmx_1) F^*(\bmz_2)
    \right\}.
\end{align}
This represents the \textit{best possible process outcome} achievable by following an optimal process trajectory from the specified initial state. Recurrent functional equations of this form are generally referred to as the Bellman equation in control theory~\cite{bertsekas2012dynamic}.

\subsection{Deriving a Tractable Optimization Problem}

The definition in Eq.~\eqref{eq:optimal_expected_cumulative_loss_def1} involves optimization over downstream trajectories. Fortunately, we can derive a tractable alternative that, perhaps unexpectedly, solves both the regression and trajectory optimization problems simultaneously.  

One of the challenges with Eq.~\eqref{eq:optimal_expected_cumulative_loss_def1} is how to handle the nested min operator. We first note that
\begin{align}
    F^*(\bmz) \leq  C(\bmz) + \sum_{\bmz'}p(\bmz' \mid \bmz,\bmx)
    F^*(\bmz')
\end{align}
holds in a transition $(\bmz,\bmx)\to \bmz'$ from any $(\bmz,\bmx)$. Assuming deterministic transitions, we approximate $F^*(\bmz)$ using a parametric model $F^\theta(\bmz)$, where $\theta$ is a set of model parameters to be learned from the training data $\calD$. Under the deterministic setting, the above inequality becomes:
\begin{align}\label{eq:determinisitic_Bellman}
    F^\theta(\bmz) \leq C(\bmz) + F^\theta(\bmz'),
\end{align}
where $\bmz'$ is the next state resulting from applying process $\bmx$ to state $\bmz$. The tightest fit is achieved by maximizing $F^\theta(\bmz)$, where the optimal $\theta$ may vary across the state $\bmz$. To find the best fit overall, we seek $\theta$ that maximizes the expected value of $F^\theta(\cdot)$ under the constraint~\eqref{eq:determinisitic_Bellman}:
\begin{align}\label{eq:defect_terminal_loss_problem}
    \max_{\theta} \sum_{\bmz} \rho(\bmz) F^\theta(\bmz) \quad 
    \text{s.t.} \quad F^\theta(\bmz) \leq C(\bmz) + F^\theta(\bmz'), \quad \forall (\bmz \to \bmz'),
\end{align}
where $\rho(\bmz)$ denotes the empirical distribution of states in $\calD$. This approach is based on the linear Bellman inequality formulation introduced in~\cite{de2003linear}, and was discussed in the present context in~\cite{Miyaguchi25advantage_RL} for the first time.

\subsection{Solving the Optimization Problem}

Incorporating Eq.~\eqref{eq:F=0_at_absorbed state}, the constraint can be rewritten as:
\begin{align}
    F^\theta(\bmz) \leq C(\bmz) + F^\theta(\bmz') (1 - \mathbb{I}(\bmz')),
\end{align}
where $\mathbb{I}(\bmz')$ is an indicator that equals 1 if $\bmz' = \bms_\perp$, and 0 otherwise. This constraint can be incorporated into the objective as the TD (time difference)-style penalty:
\begin{align}
    \max_{\theta} R(\theta \mid \mu), \quad 
    R(\theta \mid \mu) = \sum_{\bmz} \left[
        \mu \rho(\bmz) F^\theta(\bmz) 
        - \frac{1}{2}
        \left\{ 
        F^\theta(\bmz) - C(\bmz) - F^\theta(\bmz')(1 - \mathbb{I}(\bmz'))
        \right\}^2
    \right],
\end{align}
where $\mu^{-1}$ is a regularization hyperparameter. Exploiting the fact that the second term in the parenthesis can be written as 
\begin{align}
    \frac{1}{2}\{ \ldots\}^2
    &=\left\{(1-\mathbb{I}(\bmz'))[F^\theta(\bmz)-C(\bmz)-F^\theta(\bmz') ] + \mathbb{I}(\bmz')[F^\theta(\bmz)-C(\bmz)]\right\}^2
    \\
    &= \left\{(1-\mathbb{I}(\bmz'))[F^\theta(\bmz)-F^\theta(\bmz') ] + \mathbb{I}(\bmz')[F^\theta(\bmz)-y(\bmz)]\right\}^2
\end{align}
and that $\mathbb{I}(\bmz') (1 - \mathbb{I}(\bmz')) =0$ always holds, the final objective to be maximized is:
\begin{align}\label{eq:defect_terminal_loss_problem_objective}
    R(\theta \mid \mu) = 
    \frac{1}{N} \sum_{n=1}^N \left[
        \frac{\mu}{ L^{(n)}} \sum_{t=1}^{L^{(n)}} F^\theta(\bmz_t^{(n)})
        - \frac{1}{2} \left\{ y^{(n)} - F^\theta(\bmz_{L^{(n)}}^{(n)}) \right\}^2
        - \frac{1}{2} \sum_{t=1}^{L^{(n)} - 1}\!\!\!\!\mu_i \left\{ F^\theta(\bmz_{t+1}^{(n)}) - F^\theta(\bmz_t^{(n)}) \right\}^2
    \right],
\end{align}
where $\mu_i$ is a hyperparameter that can be adjusted according to data quality. More detailed mathematical analysis shows that the solution must satisfy $F^\theta(\bmz_{t+1}) \geq F^\theta(\bmz_t)$. To ensure this, we employ a positive output neural network for the difference:
\begin{align}\label{eq:G=F-F}
    G^\theta(\bmz_t, \bmz_{t+1}) \triangleq F^\theta(\bmz_{t+1}) - F^\theta(\bmz_t) = \text{ReLU}_\theta(\bmz_t \oplus \bmz_{t+1}),
\end{align}
where $\oplus$ denotes vector concatenation and $\text{ReLU}_\theta$ is the rectified linear unit activation function with a neural network parameter set $\theta$. Equations~\eqref{eq:defect_terminal_loss_problem_objective} and ~\eqref{eq:G=F-F} define the main optimization problem in the proposed PLA framework.

\subsection{Process Attribution with PLA Framework}

The PLA framework defined by Eqs.~\eqref{eq:defect_terminal_loss_problem_objective} and~\eqref{eq:G=F-F} offers several unique advantages.

First, it directly provides the attribution score:
\begin{align}\label{eq:PLA_attribution}
    \alpha_k(\xi) \triangleq G^\theta(\bmz_{k-1}, \bmz_k) \quad \text{(PLA)},
\end{align}
which is guaranteed to be non-negative. Once we learned the model parameter $\theta$, this formula can be used for any process route instance $\xi$. 

Second, the attribution score is derived from the comparison of best possible downstream outcomes with and without the process of interest, eliminating the arbitrariness of zeroing out. As shown in the next section (Fig.~\ref{fig:CT_scoring}), this allows providing informative signals to wafers whose defect density is close to the population mean. 

Finally, the term $\left\{ y^{(n)} - F^\theta(\bmz_{L^{(n)}}^{(n)})\right\}^2$ in Eq.~\eqref{eq:defect_terminal_loss_problem_objective} ensures that $F^\theta(\bmz)$ is a reasonable estimator of terminal loss at $\bmz \in \calS_T$, allowing the PLA framework to solve not only the downstream path optimization problem but also trajectory regression simultaneously.

\section{Empirical Evaluation}
\label{sec:experiments}

We applied the PLA framework to conduct root-cause analysis for a specific defect type in a state-of-the-art FEOL process. The training dataset $\calD$ was collected at the NY CREATES Albany NanoTech fab, which consists of process histories for $N=787$ wafers, covering hundreds of process steps, along with corresponding defect density measurements obtained at a process-limited yield (PLY) evaluation point. We implemented PLA in Python, where PyTorch~2.3.1 was used for the attribution function in Eq.~\eqref{eq:PLA_attribution}.

\subsection{Process and Route Embedding}

One of the critical challenges in cross-process defect attribution is the limited sample size relative to the combinatorial complexity of processing routes. Process and route embedding aims to leverage partial commonalities among routes to enable more robust prediction.

\begin{figure}
    \centering
    \includegraphics[clip, trim=0cm 2cm 0cm 1cm, width=0.49\linewidth]{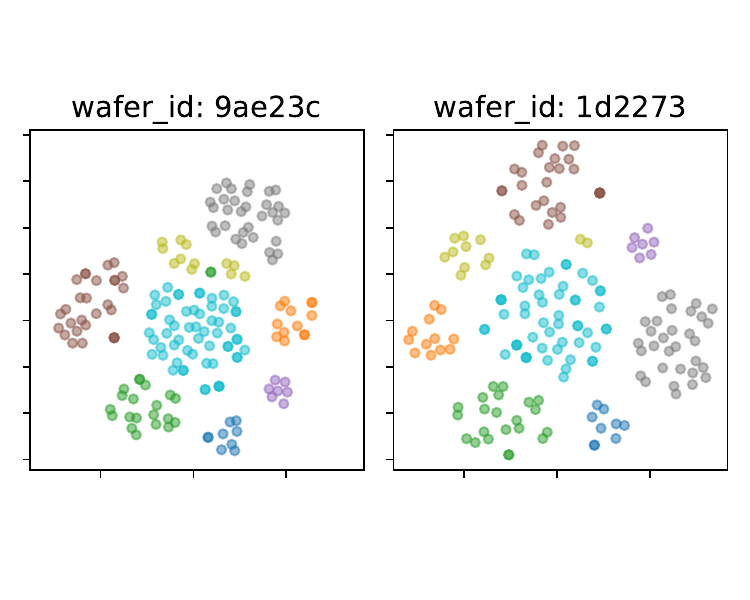}
    \hspace{1cm}
    \includegraphics[clip, trim=0cm 0cm 0cm 0cm, width=0.25\linewidth]{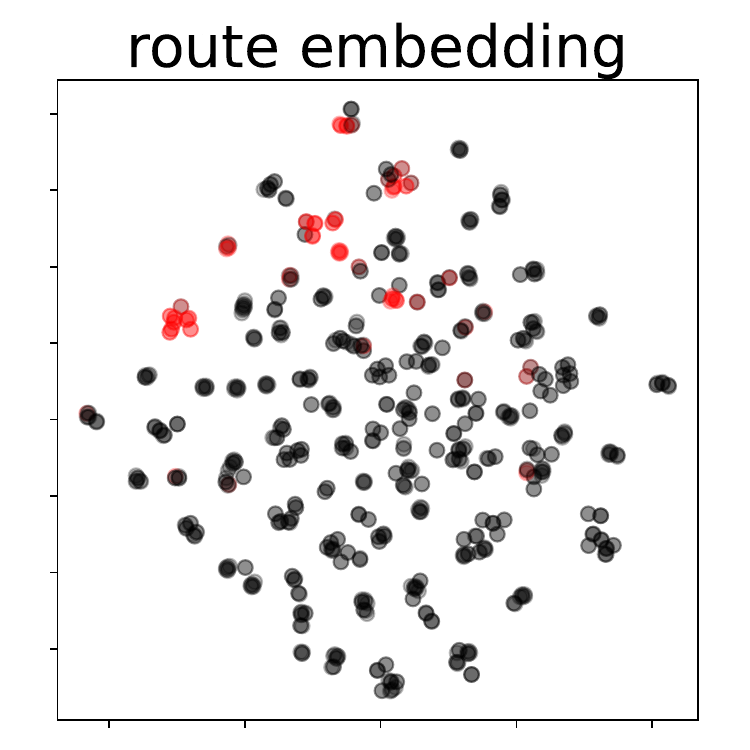}
    \caption{Process and route embeddings. \textbf{Left}: Distribution of process embeddings $\{{\bmx_k}\}_{k=1}^{L}$ for two wafer instances. Each process belongs to one of eight process types (wet process, rapid thermal processing, inspection, lithography, reactive ion etching, ion implantation, furnace, chemical mechanical polishing), which are color-coded. \textbf{Right}: Distribution of route embeddings $\{\bmz_{L^{(n)}}^{(n)}\}_{n=1}^N$ in the training dataset $\calD$, with the top 10\% highest-defect-density routes highlighted in red.}
    \label{fig:MDS}
\end{figure}

Figure~\ref{fig:MDS}~(a) shows the scatter plot of process embeddings for two wafer examples, with each point corresponding to a process step along the route. The visualization was generated using $t$-SNE as implemented in scikit-learn~\cite{van2008visualizing} with perplexity 30. Clear clustering structures are observed, suggesting that the string kernel embedding captures similarities between processing conditions effectively. Points located in close proximity typically correspond to slightly different recipe versions applied on the same tool type.

Figure~\ref{fig:MDS}~(b) shows route-level embeddings for all $N = 787$ wafers, where each point represents the entire process history of a wafer. The figure exhibits numerous micro-clusters, many of which may originate from lot-based processing patterns. Routes in the top 10\% of defect density are highlighted in red. Notable clusters in the upper region of the plot indicate the potential presence of systematic issues.

\subsection{Process Attribution}


Next, we compared the proposed PLA framework with PTR. As a baseline, we used a linear model in Eq.~\eqref{eq:regression_MLP}, which yielded a moderate cross-validated correlation of 0.61 between predicted and ground truth log defect density. In contrast, PLA, using a two-hidden-layer neural network in Eq.~\eqref{eq:G=F-F}, achieved a correlation efficient of 0.87, demonstrating a significantly improved capability for modeling the process outcomes.

\begin{figure}
    \centering
    \includegraphics[clip, trim=8.5cm 0cm 0cm 0cm, width=0.7\textwidth]{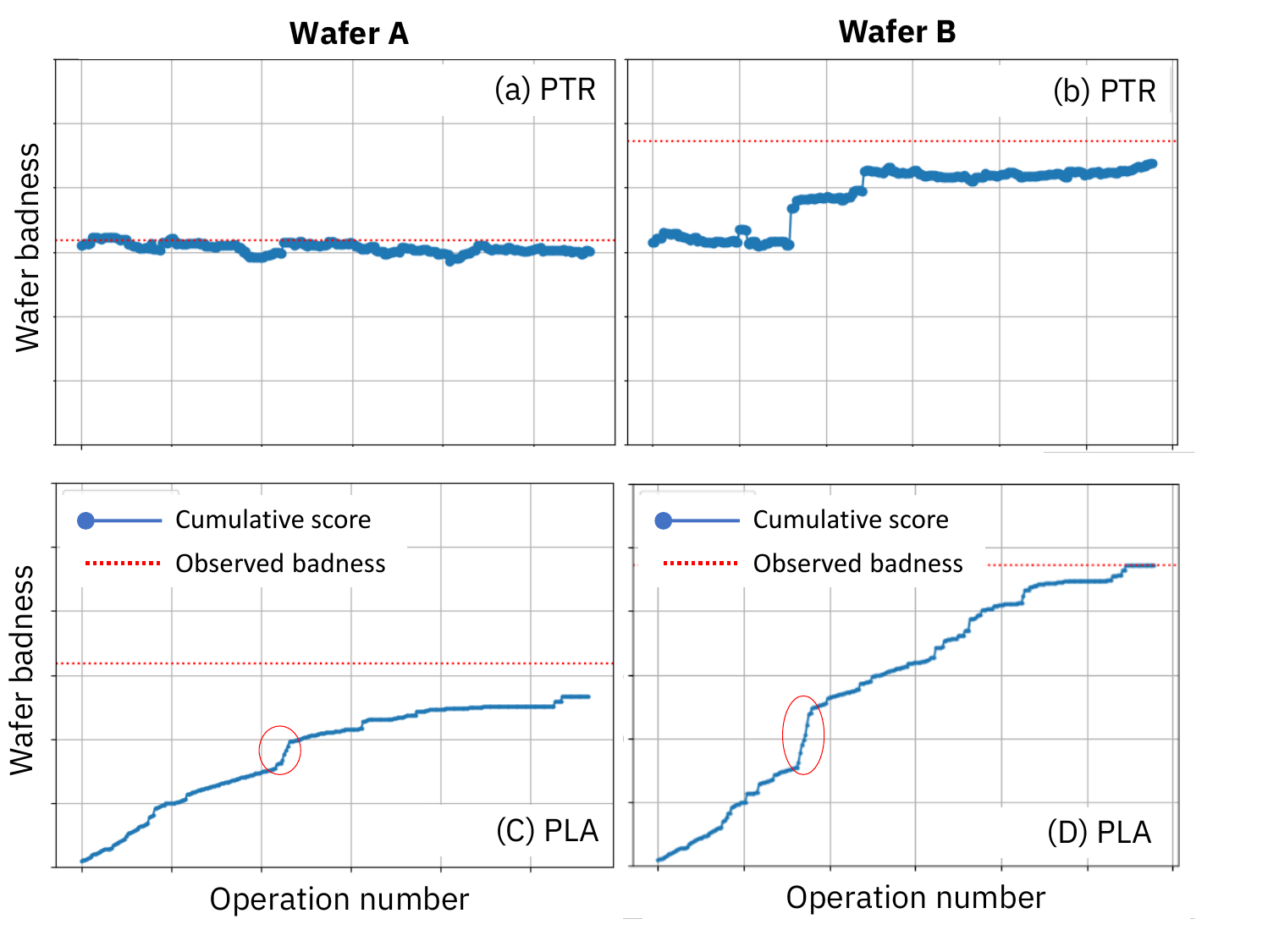}
    \caption{Comparison of PTR and PLA in terms of cumulative attribution scores. For Wafer A, the PTR-based score in (a) fails to provide informative signals, while the PLA result in (c) clearly shows how the wafer defectiveness accumulates along the route. For Wafer B, the negative attribution values in (b) complicate interpretation, whereas PLA yields consistently interpretable scores thanks to the guaranteed non-negativity.     }
    \label{fig:CT_scoring}
\end{figure}

Figure~\ref{fig:CT_scoring} compares cumulative attribution scores between PTR and PLA for a wafer from the held-out test set. Specifically, it plots the following quantities at each timestamp $\tau$ such that $t_k \leq \tau$:
\begin{align}
    \sum_{i=1}^k \alpha_i + f(\bmz_0) \quad (\text{PTR}) \quad\text{or}\quad \sum_{i=1}^k \alpha_i + F^\theta(\bmz_0) \quad (\text{PLA}).
\end{align}
For PTR, the cumulative scores at the initial and terminal processes correspond to $f(\bmz_1)$ and $f(\bmz_L)$, respectively. Similarly, for PLA, they correspond to $F^\theta(\bmz_1)$ and $F^\theta(\bmz_L)$. This visualization illustrates how the predicted defect density, referred to as `badness' in the figure, accumulates throughout the process sequence.

PTR-based attribution exhibits ups and downs due to the absence of a monotonicity constraint. This is particularly problematic for wafers with near-average defect density in Fig.~\ref{fig:CT_scoring}~(a), as the score is computed relative to the population mean due to its linear construction. In contrast, Figs.~\ref{fig:CT_scoring}~(c) and (d) show that PLA produces more stable and interpretable attribution curves.

In Fig.~\ref{fig:CT_scoring}~(c) and (d), the marked upward jumps in the attribution curves were found to correspond to unusually long wait times at specific tools, the latter of which was suspected to be the main contributor to the defect type of interest. This result highlights how PLA can effectively pinpoint problematic processes and provide actionable insights for root cause analysis.

\section{Conclusion}

We have proposed a new cross-process defect attribution framework, Potential Loss Analysis (PLA). The PLA framework addresses a fundamental challenge in sequential manufacturing: how to evaluate the responsibility of each process along the processing route.

We formalized the attribution task as a comparison of best-possible outcomes across two counterfactual partial trajectories. To the best of our knowledge, this is the first work to show that the problem can be reduced to solving a Bellman equation, enabling simultaneous regression and attribution.

Empirical evaluation on a state-of-the-art FEOL process demonstrated that PLA overcomes critical limitations of the prior partial trajectory regression approach. PLA not only improves prediction accuracy but also yields more reliable and interpretable attribution scores, making it a promising tool for data-driven, cross-process root cause analysis in semiconductor manufacturing.

\section*{Acknowledgement}

The authors gratefully acknowledge the support of NY CREATES and the Albany NanoTech Complex for providing access to state-of-the-art fabrication and characterization resources. They also extend their gratitude to Rebekah Sheraw, Monirul Islam, and Ishtiaq Ahsan for providing the PLY data and their valuable support throughout the project.

\bibliographystyle{wsc}
\bibliography{ref_ide_miya}
\end{document}